\documentclass[a4paper,11pt]{article}
\pdfoutput=1 

\usepackage{jheppub} 
\usepackage{tikz}
\usepackage[T1]{fontenc} 
\usetikzlibrary{shapes.misc}
\usetikzlibrary{decorations.pathreplacing,decorations.markings}

\tikzset{cross/.style={cross out, draw=black, minimum size=2*(#1-\pgflinewidth), inner sep=0pt, outer sep=0pt},
cross/.default={1pt}}
\tikzset{
  on each segment/.style={
    decorate,
    decoration={
      show path construction,
      moveto code={},
      lineto code={
        \path [#1]
        (\tikzinputsegmentfirst) -- (\tikzinputsegmentlast);
      },
      curveto code={
        \path [#1] (\tikzinputsegmentfirst)
        .. controls
        (\tikzinputsegmentsupporta) and (\tikzinputsegmentsupportb)
        ..
        (\tikzinputsegmentlast);
      },
      closepath code={
        \path [#1]
        (\tikzinputsegmentfirst) -- (\tikzinputsegmentlast);
      },
    },
  },
  mid arrow/.style={postaction={decorate,decoration={
        markings,
        mark=at position .5 with {\arrow[#1]{stealth}}
      }}},
}
\preprint{CALT 2020-043, IPMU 20-0109}

\title{\boldmath Singularities of thermal correlators at strong coupling}

\author[a]{Matthew Dodelson}
\author[a,b]{and Hirosi Ooguri}

\affiliation[a]{Kavli Institute for the Physics and Mathematics of the Universe,\\University of Tokyo, Kashiwa, 277-8583, Japan}
\affiliation[b]{Walter Burke Institute for Theoretical Physics,\\
California Institute of Technology, Pasadena, CA 91125, USA}

\emailAdd{matthew.dodelson\_at\_ipmu.jp}
\emailAdd{ooguri\_at\_caltech.edu}

\abstract{We analyze the singularities of the two-point function in a conformal field theory at finite temperature. In a free theory, the only singularity  
is along the boundary light cone. 
In the holographic limit, a new class of singularities emerges since 
two boundary points can be connected by a nontrivial null geodesic in the bulk,
encircling the photon sphere of the black hole. 
We show that these new singularities are resolved by tidal effects due to the black hole curvature,
by solving the string worldsheet theory in the Penrose limit. 
Singularities in the asymptotically flat black hole geometry are
also discussed.}

\begin{document} 
\maketitle
\flushbottom
\section{Introduction}
Singularities of scattering amplitudes play a fundamental role in quantum field theory. Simple poles in the kinematic invariants signify the presence of an on-shell state, and therefore contain valuable information about the theory. A less well-understood question is the role of singularities of correlation functions 
in conformal field theory (CFT) in Lorentzian position space. There are some known results. 
For instance, there is the bulk-point limit $z=\overline{z}$, where $z$ and $\overline{z}$ are the conformal cross-ratios \cite{ggp}. In
two-dimensional CFT, the four point function cannot have a singularity at
 $z=\overline{z}$, and the bulk-point singularity has to be resolved 
\cite{bulkpoint}. However,
this has not been generalized to CFTs in higher dimensions or non-conformal theories. Another known result is that the perturbative singularities are classified in terms of Landau diagrams, but this analysis does not apply to potential singularities arising from nonperturbative effects.\\
\indent In this paper we will turn our attention to singularities at finite temperature.
There, there are interesting questions even for the two-point function. We consider a conformal theory on $S^1_\beta\times S^{d-1}$ in the holographic limit, so that the theory can be analyzed via an $\text{AdS}_{d+1}$-Schwarschild black hole\footnote{The recent work \cite{alday} analyzes the opposite regime, where the ensemble is dominated by thermal AdS.}. Through the AdS/CFT duality, any null geodesic connecting two boundary points leads to a singularity in the two-point function at those points \cite{bulkcone}. This allows us to classify the singularities of the correlation function for a local bulk theory, as we will see in Sections \ref{s2} and \ref{s3}. \\
\indent Although bulk locality is a good approximation in most kinematic regimes of the correlator, there is no guarantee that stringy corrections to the propagator are small when two-points are almost null separated. Some useful intuition comes from recalling the situation near the bulk point singularity, where the legs of the bulk Landau diagram become almost lightlike. As the bulk point limit is approached, stringy corrections become more and more important, and in fact resolve the bulk point singularity. We will see that a similar effect occurs for the thermal two-point function, so that the local bulk approximation breaks down near the light cone. Whereas the bulk point singularity is resolved by the Gross-Mende effect \cite{Gross} as shown
in \cite{bulkpoint, mdooguri}, we will find that the singularities in the thermal two-point function are
resolved by worldsheet particle production.
\\
\indent The task of computing the $\alpha'$ corrections to the propagator is greatly simplified by the fact that the two points are almost null separated. In this regime we may take the Penrose limit, where the theory becomes solvable, as reviewed in Section 4. The Penrose limit captures the effects of tidal forces on strings, and the corrections to the propagator can be interpreted in terms of particle production on the worldsheet. In Sections
5 and 6, we study effects of the tidal forces on the bulk-to-bulk propagator, which
can be defined in string theory as in \cite{offshell}.
In Section 5 we show how tidal effects resolve the light-cone singularity in the bulk-to-bulk propagator at early times, when the bulk geodesic is far away from the black hole. Then in Section 6 we do a similar calculation at late times, when the geodesic wraps the photon sphere many times.\\
\indent  Once we have shown that the singularity is resolved in the bulk-to-bulk propagator, we must then argue that the same is true for the boundary two point function. As discussed in Section 7, this introduces an additional layer of complication, and requires analytic continuation of the correlation function to complex position space. Finally, in Section 8 we discuss generalizations to asymptotically flat holes.
\section{The light cone of the AdS black hole}\label{s2}
\indent In this section we will review the kinematics of null geodesics in the AdS black hole. 
Since we are interested in geodesics connecting two points on the boundary, 
they can never go inside the photon sphere. 
Using these geodesics we are able to find the location of the new
singularities on the boundary. Some of these singularities were noted in \cite{bulkcone} (see also \cite{twosided,festuccia,hoyos} for the two-sided case).
We will generalize them here and show in later sections
how they are resolved by stringy effects. \\
\indent The AdS$_{d+1}$-Schwarzschild metric is
\begin{align} 
ds^2=-\left(r^2+1-\frac{w_d  M}{r^{d-2}}\right)dt^2+\frac{dr^2}{r^2+1-\frac{  w_dM}{r^{d-2}}}+r^2d\Omega_{d-1}^2,
\end{align} 
where 
\begin{align}
w_d =\frac{8G_N\Gamma(d/2)}{(d-1)\pi^{d/2-1}}.
\end{align}
From now on we choose the normalization of $G_N$ such that $ w_d=1$. \\
\indent We consider a geodesic on the equatorial plane. The energy and angular momentum are 
\begin{align}
E=\left(r^2+1-\frac{ M}{r^{d-2}}\right)\dot{t},\hspace{10 mm}L=r^2 \dot{\phi}.
\end{align}
Using this we find 
\begin{align}
\frac{1}{2}E^2=\frac{1}{2}\dot{r}^2+ V(r),
\end{align}
where the effective potential is
\begin{align}
V(r)=\frac{L^2}{2}\left(1+\frac{1}{r^2}-\frac{ M}{r^{d}}\right).
\end{align}
Clearly for $d=2$ there is no minimum of $V$. This implies that the only boundary singularity in $d=2$ is on the ordinary light-cone.  Solving $V'=0$ for $d>2$ gives a photon sphere at 
\begin{align}
r_\gamma=\left(\frac{d  M}{2}\right)^{1/(d-2)}.
\end{align}
So for any $d>2$ there are null geodesics that come in from the boundary and escape back to infinity.\\
\indent Now let us solve for the geodesic equations. Converting $\tau$ derivatives into $r$ derivatives gives 
\begin{align}
\frac{d\phi}{dr}&=\pm\frac{1}{\sqrt{((E/L)^2-1)r^4-r^2+  Mr^{4-d}}}\notag\\
\frac{dt}{dr}&=\pm \frac{E}{L\sqrt{((E/L)^2-1)r^4-r^2+  Mr^{4-d}}\left(1+\frac{1}{r^2}-\frac{  M}{r^{d}}\right)}.
\end{align}
We are interested in the total elapsed $\Delta \phi$ and $\Delta t$. We can solve for this by first finding the turning point outside the photon sphere, and then doubling the contribution from this turning point to infinity. The turning points are at $E^2=2V$. From now on we work in $d=4$. Then we get a simple quartic equation. The roots are at $\pm r_+, \pm r_-$, where
\begin{align}
r_{\pm}=\sqrt{\frac{1\pm \sqrt{1-4    M\left((E/L)^2-1\right)}}{2((E/L)^2-1)}}.
\end{align}
Note that
\begin{align}
(r_+r_-)^2=\frac{   M}{(E/L)^2-1}.
\end{align}
Also recall that the horizon radius is 
\begin{align}
r_s=\sqrt{\frac{\sqrt{1+4   M}-1}{2}}.
\end{align}
For small black holes this is $r_s=r_\gamma/\sqrt{2}$. For large black holes the horizon is at a much smaller radius than the photon sphere, $r_s=\sqrt{r_\gamma/\sqrt{2}}$. 
  \begin{figure}[t]
        \center{\includegraphics[scale=.7]
        {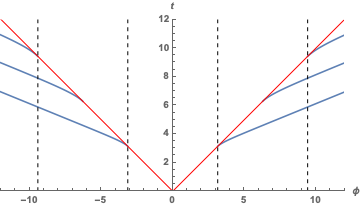}\\\includegraphics[scale=.7]
        {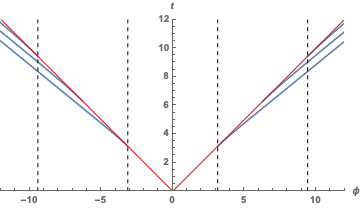}}
        \caption{\label{lightconeplot}The locations of the singularities on the boundary. The top figure is for small masses, $4   M\sim  .2$, and the bottom figure is for larger masses, $4    M\sim 2$. The angle $\phi$ has period $2\pi$, and the dashed lines are identified with each other. The standard light cone is in red, and the new singularities are in blue. }
      \end{figure}
      
  \begin{figure}[t]
        \center{\includegraphics[scale=.7]{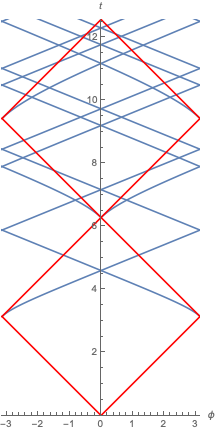}}
        \caption{\label{causticplot}The intersection pattern of the singularity curves. We see that the number of curves intersecting a given time slice grows linearly with time. Whenever two curves intersect, there is a caustic.}
      \end{figure}

      \indent The total $\Delta\phi$ can be evaluated in terms of elliptic integrals. We will take $L>0$ (the $L<0$ case can be treated in the same way). Defining $r=r_+/x$,
\begin{align}
\Delta \phi&=\frac{2r_-}{\sqrt{  M}}\int_{0}^{1}dx\, \frac{1}{\sqrt{(1-x^2)(1-(r_-/r_+)^2x^2)}}\notag\\
&=\frac{2r_-}{\sqrt{  M}}K\left(\frac{r_-^2}{r_+^2}\right).\label{elliptick}
\end{align}Let us check some limits. When $(E/L)^2-1$ is much smaller than $1/M$, $r_-\sim \sqrt{  M}$ and $r_+\sim \sqrt{1/((E/L)^2-1)}$. So 
\begin{align}
\Delta \phi =\pi+\frac{3\pi    M}{2}(E/L-1)+O((E/L-1)^2).
\label{earlytimephi}
\end{align}
This is almost equal to $\pi$, which is the answer for geodesics in pure AdS. This was to be expected since in this limit the geodesic is far away from the black hole. In the opposite limit, when we approach $E/L=\sqrt{1+1/(4   M)}$, $r_-$ and $r_+$ both approach the photon sphere. Expanding the $K$ functions, we get 
\begin{align}
\Delta \phi =-\sqrt{2}\log\left(r_+/r_--1\right).
\end{align}
\indent Now let us compute $\Delta t$. Defining $r=r_+/x$,
\begin{align}
\Delta t&=\frac{2 Er_-}{L\sqrt{  M}}\int_{0}^{1}dx\, \frac{1}{\sqrt{(1-x^2)(1-(r_-/r_+)^2x^2)}\left(1+\frac{x^2}{r_+^2}-\frac{   Mx^4}{r_+^4}\right)}\notag\\
&=\frac{2 Er_-}{L\sqrt{  M}}\frac{r_s^2\Pi\left(\frac{r_s^2}{r_+^2},\frac{r_-^2}{r_+^2}\right)+(1+r_s^2)\Pi\left(-\frac{1+r_s^2}{r_+^2},\frac{r_-^2}{r_+^2}\right)}{1+2r_s^2}.
\end{align}
In the limit of an infinitely large black hole, we get $\Delta \phi=\pm \Delta t$. For a black hole with finite mass, the resulting boundary singularities are shown in blue curves in
Figure \ref{lightconeplot}. Note that the blue curve first appears
at $\Delta \phi = \pm \pi$ since this is where the non-trivial bulk null geodesics start deviating from the boundary light-cone as in (\ref{earlytimephi}).  Figure \ref{causticplot}
takes into account the $2\pi$ periodicity in $\phi$.

\indent The lowest blue curve in Figure \ref{lightconeplot} was noted in \cite{bulkcone}. 
There are other blue curves since null geodesics can hit the
boundary, bounce back into the bulk, and escape back to the boundary again, and this can be repeated many times. This leads to more singularities in the correlation function, at $(\phi,t)=(n\Delta \phi,n\Delta t)$ for any integer $n>0$. The full light cone is depicted in Figure \ref{lightconeplot}. Note that the singularity curves can intersect each other, leading to caustics where more than one null geodesic connects two boundary points. These
singularity curves become increasingly dense as $t$ increases. 
This is shown in Figure \ref{causticplot}.\\
\indent 
At late times, the blue curves approach a straight line, with slope
\begin{align}
v_\gamma\equiv \lim_{\Delta t \rightarrow \infty} \frac{|\Delta \phi|}{\Delta t}=\sqrt{1+\frac{1}{4  M}}.
\label{PhotonSphereVelocity}
\end{align}
We can understand this as follows. At late times the geodesic spends most of its time near the photon sphere. So the effective velocity is $\sqrt{g_{tt}/g_{\phi\phi}}$ evaluated at the photon sphere, which indeed gives $\sqrt{1+1/(4   M)}$.\\
      \section{Computing the correlator near the singularity}\label{s3}
      In the last section we found the location of the singularity. Now we want to compute the behavior of the correlation function as we approach the singularity. We will work in the geodesic approximation, which relies on the dimension of the external operators being very large.

\subsection{The geodesic approximation}
For large masses, we need to consider geodesics that are slightly spacelike, and then take the limit as they become light-like. The geodesic approximation to the propagator is then given by $e^{-ml}$, where $l$ is the proper length of the geodesic. For spacelike geodesics the potential is modified to 
\begin{align}
V(r)=\frac{1}{2}\left(L^2-r^2\right)\left(1+\frac{1}{r^2}-\frac{  M}{r^4}\right).
\end{align}
There are now new turning points at large imaginary $r$. 
The imaginary turning points can be found by expanding around large $r$. They are at
\begin{align}
r_{\text{im}}^2\approx -(E^2-L^2)
\end{align}
So therefore we can write 
\begin{align}
E^2-2V=(r^2-r_{\text{im}}^2)(1-r_+^2/r^2)(1-r_-^2/r^2).
\end{align}
\indent Now we can compute the proper length $l$. Putting in a cutoff at $r_{\text{max}}$,
\begin{align}
l=2\int_{r_+}^{r_{\text{max}}}\frac{dr}{\dot{r}}=2\int_{r_+}^{r_{\text{max}}}\frac{dr}{\sqrt{(r^2+|r_{\text{im}}|^2)(1-r_+^2/r^2)(1-r_-^2/r^2)}}.
\end{align}
This is clearly log divergent at large $r$. To do the integral, let us separate it into two parts. The first part is from $r_+$ to $r_0$ with $|r_{\text{im}}|\gg r_0\gg r_+$, and the second part is from $r_0$ to $r_{\text{max}}$. In the first part of the integral, the integrand is suppressed by $1/|r_{\text{im}}|$, so we can ignore it near the light-cone. The second part of the integral is 
\begin{align}
2\int_{r_0}^{r_{\text{max}}}\frac{dr}{\sqrt{r^2+|r_{\text{im}}|^2}}\approx 2\log\left(\frac{r_{\text{max}}}{|r_{\text{im}}|}\right).
\end{align}
The boundary correlator is obtained by exponentiating the renormalized length,
\begin{align}
e^{-ml_{\text{ren}}}=\left(E^2-L^2\right)^{m}
\end{align}
\subsection{Converting $E$ and $L$ to boundary variables}
We now need to trade $E$ and $L$ for boundary variables $\Delta \phi,\Delta t$. To do this we have integrals of the form
\begin{align}
\Delta \phi &=2L\int_{r_+}^{\infty}\frac{dr}{\sqrt{(r^2+|r_{\text{im}}|^2)(r^2-r_+^2)(r^2-r_-^2)}}\notag\\
\Delta t&=2E\int_{r_+}^{\infty}\frac{dr}{\left(1+\frac{1}{r^2}-\frac{   M}{r^4}\right)\sqrt{(r^2+|r_{\text{im}}|^2)(r^2-r_+^2)(r^2-r_-^2)}}.
\end{align}
Let us start with $\Delta \phi$. Since $r>r_{\pm}$ in the integration region, we can Taylor expand in $r_{\pm}/r$. This gives integrals of the form
\begin{align}
f(r_+,r_-)L\int_{r_+}^{\infty}\frac{dr}{\sqrt{r^2+|r_{\text{im}}|^2}r^{2n}},\hspace{10 mm}n\ge 1.
\end{align}
We want to keep terms up to order $1/r^2_{\text{im}}$. The contribution of the lower endpoint to the integral contains terms of order $1/r_{\text{im}}$ for all $n$. This is just the answer for null geodesics. The only term of order $1/r_{\text{im}}^2$ comes from the upper endpoint for $n=1$. So we get (again taking $L>0$),
\begin{align}
\Delta \phi =\Delta \phi_\text{null}(E/L)-\frac{2L}{E^2-L^2},
\end{align}
where $\Delta \phi_{\text{null}}$ is defined by (\ref{elliptick}). Similarly 
\begin{align}
\Delta t=\Delta t_\text{null}(E/L)-\frac{2E}{E^2-L^2}.\label{deltat}
\end{align}
\indent We now need to solve for $E$ and $L$. This is in general complicated, but simplifies in several limits. For example let us consider the late time limit $r_+\to r_-$. In this limit we have to solve 
\begin{align}
\Delta \phi &=-\frac{1}{\sqrt{2}}\log\left(\frac{E}{L}-v_\gamma\right)-\frac{2L}{E^2-L^2}\notag\\
\Delta t&=-\frac{1}{\sqrt{2}v_\gamma}\log\left(\frac{E}{L}-v_\gamma\right)-\frac{2E}{E^2-L^2}
\end{align}
 Solving for $E$ and $L$, we get
 \begin{align}
 E=\frac{2v_\gamma}{\Delta \phi -v_\gamma\Delta t},\hspace{10 mm}L=\frac{2}{ \Delta \phi -v_\gamma\Delta t }
 \end{align}
 Finally we plug into the correlation function to get 
\begin{align}
\left(  (v_\gamma \Delta t-\Delta \phi )^2\right)^{-m}.
\end{align}
This gives a singularity
at $\Delta \phi = v_\gamma \Delta t$ with the same strength as that at
the boundary light cone $\Delta \phi =  \Delta t$. The same calculation can be done for negative $\Delta \phi$ with similar results.

\section{Review of string theory in the Penrose limit}\label{s4}
In the previous section we presented some evidence that a new singularity is present at infinite 't Hooft coupling. In Appendix \ref{freefield} we show that this singularity is in fact absent at zero coupling. This suggests the possibility that the singularity is only present at infinite coupling, and is resolved at any finite coupling. Now we would like to understand what happens at large but finite coupling. To do so we need to analyze stringy corrections to the propagator. The worldsheet sigma model in the full black hole geometry is intractable, but fortunately we are only interested in the behavior of the propagator in the near vicinity of the light cone. There is a well-known procedure for studying the geometry close to a given null geodesic, which is to take the Penrose limit. The Penrose limit includes information about the tidal force near the null geodesic. In this limit string theory becomes solvable, so we can compute the propagator exactly. In this section we will briefly review the features of string theory in the Penrose limit.  More details can be found in the review article \cite{blau}. In this paper, we will discuss the case of closed strings. Generalization to
the open string case should be straightforward. 
\\
\indent In Brinkmann coordinates, the general plane wave metric is 
\begin{align}
ds^2=2du\, dv+A_{ab}(u)x^{a}x^b du^2+d\vec{x}^2.
\end{align}
Here $A_{ab}$ is a $(d-1)\times(d-1)$ dimensional matrix, where as usual $d+1$ is the dimension of spacetime. The vacuum Einstein equations require that $A_{ab}$ is traceless, which means that there is necessarily at least one negative eigenvalue and one positive eigenvalue, unless $A_{ab}$ is identically equal to zero. The worldsheet theory in this background is solved by going to light-cone gauge, $u=p_v\tau$. In this gauge the equations of motion for the transverse modes is \cite{blau,horowitzsteif}
\begin{align}
\ddot{X}^a_n=\left(p_v^2A_{ab}(p_v\tau)-n^2\delta_{ab}\right)X^b_n.
\end{align}
Therefore we just have a collection of coupled harmonic oscillators with a time-dependent frequency matrix. It follows that the theory can be analyzed using the standard techniques of time-dependent quantum mechanics. \\
\indent Let us recall the simplest examples. First, the Penrose limit of AdS or flat space for any null geodesic is flat space, $A_{ab}=0$. The interpretation of this statement is that the tidal force is equal to zero. A more nontrivial example is $\text{AdS}_5\times S^5$, where the null geodesic is a great circle on the $S^5$. This corresponds to the plane wave limit of AdS/CFT \cite{bmn}. In this case the matrix $A_{ab}$ is constant. \\
\indent Now let us turn to the case of interest. For $\text{AdS}_5$-Schwarschild, the plane wave matrix is \cite{blau}
\begin{align}
A_{11}=\frac{4L^2    M}{r^6}=-2A_{22}=-2A_{33}.
\end{align}
In particular, the equations of motion are diagonal,
\begin{align}
\ddot{X}^a_n=-(\omega^a_n)^2X^a_n.\label{EOM}
\end{align}
Here the frequencies $\omega^a_n$ are defined by
\begin{align}
(\omega^1_n)^2&=n^2-\frac{4p_v^2L^2M}{r(p_v\tau)^6}\notag\\
(\omega^{2,3}_n)^2&=n^2+\frac{2p_v^2L^2M}{r(p_v\tau)^6}.\label{frequencies}
\end{align}
\indent Our goal is to compute the bulk-to-bulk propagator in the Penrose limit, and to show that it is nonsingular on the light cone. Unfortunately, the equations of motion for $\ddot{X}^a_n$ are analytically intractable, so we will need to resort to several approximation schemes to solve them. The three relevant approximations are the Born approximation, the shockwave approximation, and the WKB approximation. When the geodesic is far away from the black hole, the Born approximation and the shockwave approximation can be combined to compute the propagator. The WKB approximation is valid in the opposite limit, when the geodesic passes very close to the photon sphere. We will analyze these limits in the next two sections. \\
\indent Before turning to the calculation, let us state our strategy for computing the propagator. The quantity of interest is a simple generalization of the flat space propagator \cite{offshell,gsw} and can be interpreted as an annulus amplitude on a pair of D$_{-1}$
branes placed at two bulk points. Expanding the string modes into the classical piece plus fluctuations, we find (disregarding an overall normalization factor)
\begin{align}
\langle p_v,x^a_f,u_f|p_v,x^a_i,u_i\rangle&=\int_{X^a(\tau_i,\sigma)=x^a_i}^{X^a(\tau_f,\sigma)=x^a_f}DX^a\,  e^{iS[X^a]}\notag\\
&=\frac{G_0(p_v,u_f,u_i,x_f^a,x_i^a)}{\prod_{a=1}^{3}\prod_{n=1}^{\infty}\det\left(-\partial_\tau^2-n^2+p_v^2A_{aa}(p_v\tau)\right)},
\end{align}
where $G_0$ is the zero mode propagator. Here the determinants are evaluated subject to the boundary conditions on the path integral. Also, at the end of the calculation we project onto the final and initial vacuum states. This involves taking $\tau_f\to \tau_f(1-i\epsilon)$ and $\tau_i\to \tau_i(1-i\epsilon)$, and then taking $\tau_f\to\infty$ and $\tau_i\to -\infty$. \\
\indent Once we compute the propagator in $p_v$ space, we can Fourier transform to position space,
\begin{align}
G(u_f,u_i,v_f-v_i,x^a_f,x^a_i)=\int_{-\infty}^{\infty} dp_v \, e^{ip_v (v_f-v_i)}\langle p_v,x^a_f,u_f|p_v,x^a_i,u_i\rangle.
\end{align}
\noindent However, if we are only interested in showing that the singularity is resolved, we can take a shortcut. The magnitude of the propagator on the light cone is bounded by the triangle inequality,
\begin{align}
|G(u_f,u_i,v_f-v_i=x_i^a=x_f^a=0)|&\le \int_{-\infty}^{\infty} dp_v\, | \langle p_v,x^a_f,u_f|p_v,x^a_i,u_i\rangle | \\
&= \int_{-\infty}^{\infty} dp_v\, |G_0(p_v,u_f,u_i,x_f^a=x_i^a=0)\langle \text{out},p_v|\text{in},p_v\rangle|\notag
.\label{bound}
\end{align}
Here $G_0$ is the zero mode propagator, and
$|\text{in},p_v\rangle$ and $|\text{out},p_v\rangle$ are the vacuum states for worldsheet oscillators 
in the far past and far future respectively. If we can show that the integral on the right hand side converges, then it follows that the left hand side is finite. Therefore in order to bound the propagator on the light cone, we only need to compute the particle production of stringy modes in the vacuum.

 \section{Bulk singularity resolution at early times}\label{s5}
 In this section we will demonstrate how the light cone singularity in the bulk to bulk propagator is resolved at early times. We are specifically interested in the propagator between two points far outside the black hole, 
 \begin{align}
 \langle \Phi(r_f,t,\phi)\Phi(r_i,0,0)\rangle,
 \end{align}
 where $r_f,r_i \gg r_\gamma $.
\subsection{Small $p_v$}\label{smallp}
For small $p_v$ the tidal forces are small and can be treated in perturbation theory. 
The tidal forces generate particle production on the worldsheet. In this case, 
the overlap between the in and out state can be computed using
 the normalization of the squeezed vacuum as
\begin{align}
|\langle \text{out},p_v|\text{in},p_v\rangle|=\prod_{a=1}^{3}\prod_{n=1}^{\infty}(1+\langle N^a_n\rangle)^{-1/2},\label{squeezed}
\end{align}
where $\langle N^a_n\rangle$ is the expectation value of the number operator in the
$n$-th excitation in the $a$-th direction. In this expression we have included the contribution of both left and right movers. The expectation value is given by \cite{horowitzsteif}
\begin{align}
\langle N^a_n\rangle=\frac{p_v^2}{4n^2}\left|\tilde{A}_{aa}\left(\frac{2n}{p_v}\right)\right|^2,
\end{align}
where $\tilde{A}_{aa}$ is the Fourier transform of $A_{aa}$. We can do this Fourier transform as follows. Since $r_-\ll r_+$, we can approximate the radial coordinate as 
\begin{align}
r=\sqrt{\frac{L^2+(E^2-L^2)^2u^2}{E^2-L^2}}.
\end{align}
Taking the $x^1$ direction as an example, we need to do the integral 
\begin{align}
\tilde{A}_{11}(k)=4L^2  M\int_{-\infty}^{\infty}\frac{du}{r(u)^6}e^{iku}=\frac{\pi   M}{2Lr_+^4}e^{-|k|r_+^2/L} \left(3L^2+3L|k|r_+^2+k^2r_+^4\right)
\end{align}
Therefore 
\begin{align}
\langle N^1_n\rangle=\left(\frac{\pi   M}{4np_vLr_+^4}\right)^2e^{-4nr_+^2/(p_vL)} \left(3p_v^2L^2+6p_vLnr_+^2+4n^2r_+^4\right)^2.
\end{align}
From the exponential factor, we see that particle production is exponentially suppressed except in the small $n$ regime,
\begin{align}
n\ll \frac{p_vL}{r_+^2},
\end{align}
where we have assumed that $p_vL\gg r_+^2$. In this limit we have
\begin{align}
\langle N^1_n\rangle=4\langle N^{2,3}_n\rangle=\left(\frac{3\pi p_vL   M}{4nr_+^4}\right)^2,
\label{born}
\end{align}
which is small for $p_vL\ll r_+^4/M$.\\
\indent Since the $\langle N^a_n\rangle$'s are small, we can approximate (\ref{squeezed}) as
\begin{align}
|\langle \text{out},p_v|\text{in},p_v\rangle| \approx 1-\frac{1}{2}
\sum_{a=1}^{3}\sum_{n=1}^{\infty}\langle N^a_n\rangle,
\end{align}
For $p_vL\ll r_+^2$ the sum is exponentially small. For $ r_+^2\ll p_v L\ll r_+^4/M  $ the sum can be performed using the approximation (\ref{born}), which gives
\begin{align}
1-3\left(\frac{3\pi p_vL  M}{8r_+^4}\right)^2\sum_{n=1}^{\infty}\frac{1}{n^2}=1-2\left(\frac{3\pi^2p_vL   M}{16r_+^4}\right)^2.
\end{align}
We see that to first order in $p_vL$, the overlap is slightly smaller than one. However this is not enough to resolve the singularity, which arises from a divergence in the Fourier transform of the propagator at large $p_vL$. We turn to this limit next.
\subsection{Large $p_v$}
For $p_vL\gg r_+^2$, the interaction only occurs over a small range of $\tau$. To see this, we write 
\begin{align}
p_v^2A_{aa}(p_v\tau)\propto \pm  \frac{  M}{(\epsilon r_+)^2} \left(\frac{\epsilon^2}{\tau^2+\epsilon^2}\right)^3,\hspace{10 mm}\epsilon=\frac{r_+^2}{p_vL}\ll 1.
\end{align}
From this equation it is evident that at large $p_vL$ the potential is localized at $\tau=0$, so it is as if the string hits a shockwave at time $\tau=0$, and propagates freely elsewhere. This approximation is valid if the modes vary slowly on the scale of $\epsilon$, so that $n\ll 1/\epsilon$. On the other hand the Born approximation of Section \ref{smallp} is valid when $n\gg M/(\epsilon r_+^2)$. It follows that the Born approximation and the shockwave approximation have an overlapping regime of validity. This is fortunate, since it implies that the calculation is under control for all values of $n$.\\
\indent We now proceed similarly to the analysis of strings propagating in a shockwave \cite{ggm} (see also \cite{divecchia} for a similar computation in a different context). We will do the calculation for $X^1_n$; the other two modes are treated in the same manner. Integrating the equation of motion (\ref{EOM}) gives a discontinuity in the first derivative of $X^{1}_n$, 
\begin{align}
\lim_{\delta\to 0}(\dot{X}^1_n(\delta)- \dot{X}^1_{n}(-\delta))&=\frac{4  M}{(\epsilon r_+)^2}X^1_{n}(0)\lim_{\epsilon\to 0}\int_{-\infty}^{\infty}d\tau \left(\frac{\epsilon^2}{\tau^2+\epsilon^2}\right)^3\notag\\
&=\frac{3\pi p_vL   M}{2 r_+^4}X^1_{n}(0).
\end{align}
We make the ansatz 
\begin{align}
X^1_{n}=
\begin{cases}
a_{1n}^\dagger e^{in\tau}+a_{1n}e^{-in\tau}\text{ for }\tau<0\\b_{1n}^\dagger e^{in\tau}+b_{1n}e^{-in\tau}\text{ for }\tau>0\label{x1ansatz}
\end{cases}
\end{align}
Assuming that $X^{1}_n$ is continuous at $\tau=0$, the solution to the differential equation is 
\begin{align}
b_{1n}&=\left(1+\frac{3\pi i    Mp_vL}{4 n r_+^4}\right)a_{1n}+\frac{3\pi i  Mp_vL}{4 nr_+^4}a_{1n}^\dagger.
\end{align}
The magnitude of the Bogoliubov coefficient $\beta^1_n$ is therefore 
\begin{align}
|\beta^1_n|=\frac{3\pi Mp_v L}{4nr_+^4},
\end{align}
and the number of produced particles is the square of $|\beta^1_n|$. This gives the same result as in the Born approximation (\ref{born}), but extrapolated to large $p_v$. Therefore at large $p_v $ there is a large range of mode numbers $n$ with a large expectation value of the number operator.\\
\indent In the shockwave approximation the magnitude of the overlap becomes 
\begin{align}
|\langle \text{out},p_v|\text{in},p_v\rangle|&=\prod_{n=1}^{\infty}\left(1+\left(\frac{3\pi p_v L  M}{4nr_+^4}\right)^2\right)^{-1/2}\left(1+\left(\frac{3\pi p_v L M}{8nr_+^4}\right)^2\right)^{-1}\notag\\
&=\frac{\left(\frac{3\pi^2 p_vLM}{4r_+^4}\right)^{3/2}}{2\sqrt{\sinh\left(\frac{3\pi^2 p_vLM}{4r_+^4}\right)}\sinh\left(\frac{3\pi^2 p_vLM}{8r_+^4}\right)}.\label{detmag}
\end{align}
At small $p_v$ this approaches unity, and at large $p_v$ it is exponentially suppressed.  \\
\indent Now that we have computed the overlap of the initial and final vacua, we may bound the propagator using (\ref{bound}). Recall that the zero mode propagator near the light cone is equal to \cite{poisson}
\begin{align}
G_0(u_f,u_i,v_f-v_i, x_f^a=x_i^a=0)=\frac{\Delta^{1/2}(r_f,r_i)}{((u_f-u_i)(v_f-v_i)+i\epsilon)^{3/2}}.
\end{align}
Here the Van-Vleck determinant $\Delta$ is independent of $v_f-v_i$, and we will not bother computing it. Fourier transforming gives
\begin{align}
G_0(p_v,u_f,u_i,x_f^a=x_i^a=0)=\frac{\Delta^{1/2}(r_f,r_i)}{(u_f-u_i)^{3/2}}\sqrt{p_v}\Theta(p_v)
\end{align}
The propagator on the light cone is therefore bounded as
\begin{align}
|G(u_f,u_i,v_f-v_i=x_i^a=x_f^a=0)|&\le  \frac{\Delta^{1/2}(r_f,r_i)}{(u_f-u_i)^{3/2}}\int_{0}^{\infty} dp_v\,\frac{\sqrt{p_v}\left(\frac{3\pi^2 p_v LM}{4r_+^4}\right)^{3/2}}{2\sqrt{\sinh\left(\frac{3\pi^2 p_v LM}{4r_+^4}\right)}\sinh\left(\frac{3\pi^2 p_v LM}{8r_+^4}\right)}\notag\\
&\sim  \Delta^{1/2}(r_f,r_i)\left(\frac{r_+^3}{M(r_f+r_i)\alpha' }\right)^{3/2}
\end{align}
In deriving this equation we used the relation $r=\sqrt{E^2-L^2}|u|$ at large $r$, and also restored a factor of the string length.  The integrand is exponentially suppressed at large $p_v$, so the integral is finite and the light cone singularity is resolved. This is analogous to the resolution of the bulk point singularity in Mellin space \cite{mdooguri}, where the divergence at large Mellin energy is cut off by stringy corrections. 
 \section{Bulk singularity resolution at late times}\label{s6}
 \indent We now turn to the late time limit, in which the geodesic wraps the photon sphere many times.\footnote{During the preparation of this manuscript we received the paper \cite{martinec}, where calculations similar to those presented
in this section appeared in a different context. }
 The geodesic stays close to the photon sphere for a long time, and is approximately circular in this region. Therefore the WKB approximation is appropriate. To show this, note that the system is adiabatic when the frequencies satisfy $\partial_\tau(\omega^a_n)^{-1}\ll 1$. For $n=0$ this becomes
 \begin{align}
 \frac{r^2\partial_u r}{L \sqrt{   M}}=\sqrt{((r/r_+)^2-1)((r/r_-)^2-1)}\ll 1.
 \end{align}
In the late time limit $r_+, r_-\approx r_\gamma$, this is valid near the photon sphere,
 \begin{align}
\frac{r-r_\gamma}{r_\gamma}\ll 1. 
 \end{align}
The same is true for the modes with $n\not=0$. \\
\indent In fact, the frequencies are not just adiabatically evolving, they are constant throughout the region near the photon sphere. Evaluating the frequencies (\ref{frequencies}) at the photon sphere, we find that the frequency for the $x^1$ direction is
\begin{align}
(\omega_n^1)^2=n^2-\frac{p_v^2 L^2 }{2M^2}.
\end{align}
A mode is unstable if the frequency is imaginary. At large $p_v$, this is true for 
\begin{align}
n<n_{\text{max}}=\frac{p_v L }{\sqrt{2}M}.
\end{align}
Therefore the number of unstable modes grows linearly with $p_v$. The $x^{2,3}$ directions are stable and we will not need to consider them here.\\
\indent Now let us consider the behavior of the solutions of the equation of motion. At large $r$ we just have a free string. As the mode propagates in time, eventually it enters the region near the photon sphere. The solution is then 
 \begin{align}
 X^1_n(u)=\exp\left(\frac{|\omega^1_n|}{p_v}\int^u du' \right).
 \end{align}
The integral in the exponent can be done by changing variables, 
 \begin{align}
\int^{u} du'&=- \frac{1}{2\sqrt{E^2-L^2}} \int^{r(u)} \frac{dr'}{\sqrt{(1-r_+/r')(1-r_-/r')}}\notag\\
&=-\frac{2\sqrt{2}M}{L}\text{ArcSinh}\left(\sqrt{\frac{r(u)/r_--1}{r_+/r_--1}}\right).
 \end{align}
 Here we have assumed that $\dot{r}<0$, and also used $r-r_\gamma\ll r_\gamma$. When $\dot{r}>0$ the answer flips signs. Therefore the outgoing mode at radius $r$ with $r/r_--1\ll r_+/r_--1$ satisfies (up to an overall constant)
 \begin{align}
 X^1_n(r,\text{out})=\left(\frac{r/r_--1}{r_+/r_--1}\right)^{\frac{2\sqrt{2}M|\omega^1_n|}{p_vL}}X^1_n(r,\text{in}).
 \end{align}
The adiabatic approximation breaks down at the end of the region near the photon sphere. At this radius we have
\begin{align}
 X^1_n(r,\text{out})\propto \frac{X^1_n(r,\text{in})}{(r_+/r_--1)^{2\sqrt{1-n^2/n_{\text{max}}^2}}}.
\end{align}
We have left out an order one constant on the right hand side, which cannot be unambiguously computed in the adiabatic approximation.\\
\indent We see that the net effect of the propagation through the adiabatic region is a large amplification factor. Assuming that we can neglect particle production in the nonadiabatic region, the Bogoliubov coefficient is equal to this factor \cite{horowitzsteif}, 
\begin{align}
|\beta^1_n|\propto (r_+/r_--1)^{-2\sqrt{1-n^2/n_{\text{max}}^2}}
\end{align}
The expectation value of the number operator is the square of this,
 \begin{align}
 \langle N^1_n\rangle\propto (r_+/r_--1)^{-4\sqrt{1-n^2/n_{\text{max}}^2}}\gg 1.
 \end{align}
Note that we have not solved the equations of motion in the nonadiabatic region, so we must assume that the main contribution to the particle production comes from the adiabatic region. Since the geodesic spends a long time near the photon sphere, we expect this to be this case, but we have not shown it explicitly.\\
 \indent Finally, since $\langle N^1_n\rangle$ is large, we may compute the overlap between the in and the out state by multiplying $\langle N_n\rangle^{-1/4}$ over all the modes,
\begin{align}
|\langle \text{out},p_v|\text{in},p_v\rangle|\propto \exp\left(2\sum_{n=1}^{n_{\text{max}}}\sqrt{1-\frac{n^2}{n^2_{\text{max}}}}\log(r_+/r_--1)\right).
\end{align}
At large values of $n_{\text{max}}$, we can approximate the sum by an integral. We get 
\begin{align}
\exp\left(\frac{\pi p_vL}{2\sqrt{2}  M}\log(r_+/r_--1)\right).
\end{align}
This is exponentially suppressed at large $p_v$, so the singularity is resolved by the same argument as at early times. 
\section{Singularity resolution in boundary correlators} \label{s7}
In the previous two sections we analyzed the bulk-to-bulk propagator. Naively this is sufficient for computing boundary correlators, since by the AdS/CFT dictionary we have 
\begin{align}
\langle O(t,\phi)O(0,0)\rangle=\lim_{r\to\infty}r^{2\Delta}\langle \Phi(t,r,\phi)\Phi(0,r,0)\rangle.
\end{align}
However, we have only computed the bulk-to-bulk propagator on the light cone, and the limits of going to the light cone and taking $r\to\infty$ do not commute. Therefore we must treat the boundary correlator separately. In this work we will only discuss the early time case.\\
\indent For boundary correlators at fixed $t$ and $\phi$, the proper distance between the two boundary points grows with the cutoff radius $r$. Therefore instead of expanding around a null geodesic, we must expand the metric around a spacelike geodesic. This is done in Appendix \ref{tidaltensor}. Once we have the metric, we can compute the propagator using the Euclidean Polyakov path integral in covariant gauge. Proceeding as in \cite{offshell}, we find
\begin{align}
\int_{0}^{\infty}\frac{d\ell}{\ell^{1/2}}\frac{\exp\left(-\frac{\Delta \tau}{2}\left(\frac{1}{\ell}+\ell m^2\right)\right)}{\prod_{a=0}^{3}\prod_{n=-\infty}^{\infty}\sqrt{\det\left(-\ell^{-2}(\partial_\tau^2-{\Phi^a}_a)+n^2\right)}},
\end{align}
where $\ell$ is the worldsheet modulus. Here we have neglected an overall $L$-independent factor.\\
\indent Actually, this is not quite correct, since we have made the implicit assumption that the path integral is convergent. This assumption is not guaranteed since there could be a negative or zero eigenvalue, corresponding to a fluctuation mode which can leave the near-geodesic region without giving a suppressed contribution to the path integral. We can understand this quantitatively by solving for a negative eigenvalue. The eigenvalue equation (say for the $x^2$ direction) is 
\begin{align}
(-\ell^{-2}(\partial_\tau^2-{\Phi^{2}}_2)+n^2-\lambda )f_\lambda=0
\end{align}
The solution that vanishes at $\tau=\tau_i$ is 
\begin{align}  
f_\lambda(\tau)=\sinh(\ell\sqrt{n^2-\lambda}(\tau-\tau_i))+\frac{3\pi LM}{4\ell \sqrt{n^2-\lambda}r_+^4}\sinh(\ell\sqrt{n^2-\lambda}\tau)\sinh(\ell\sqrt{n^2-\lambda}\tau_i)\Theta(\tau).
\end{align}
Here we have assumed that $\ell\gg 1$ so that we can neglect the constant term in the tidal tensor. Setting $\tau=\tau_f$ and taking $\tau_f,-\tau_i\to \infty$, we find 
\begin{align}
f_\lambda(\tau_f)=\frac{\exp(\ell\sqrt{n^2-\lambda}\Delta \tau)}{2}\left(1-\frac{3\pi ML}{8\ell \sqrt{n^2-\lambda}r_+^4}\right).
\end{align}
Solving for $\lambda$ then gives
\begin{align}
\lambda=n^2-\left(\frac{3\pi ML}{8\ell r_+^4}\right)^2.\label{eigenvalue}
\end{align}
Therefore a negative eigenvalue exists for all $L>8nr_+^4\ell /(3\pi M)$. \\
\indent Evidently for large enough $L$ the path integral does not converge in the Euclidean regime of real $\ell$, and the near-geodesic approximation breaks down. This means that we cannot actually compute the correlation function for real values of $t,\phi$ close to the singularity. Instead, we can approach the singularity along the imaginary $t$ axis, 
\begin{align}
\lim_{\epsilon\to 0}\langle O(\Delta t_{\text{null}}(r_+)-i\epsilon ,\Delta \phi_{\text{null}}(r_+))O(0,0)\rangle.
\end{align}
Solving (\ref{deltat}) for $L$ at large $r_+$, we find 
\begin{align}
L=\frac{2r_+^2}{\Delta t_{\text{null}}(r_+)-t}.
\end{align}
Therefore we are interested in the correlator in the limit 
\begin{align}
L=-\frac{2ir_+^2}{\epsilon}\to -i\infty.
\end{align}
In this regime there is no negative eigenvalue, as is clear from (\ref{eigenvalue}). Therefore we can use the near-geodesic approximation.
\\ \indent We can now evaluate the determinants at early times as in Appendix \ref{phase}. Note that there is a constant term in the tidal tensor (\ref{tidalequation}) proportional to $1/\ell_{\text{AdS}}^2$. If we assume that the integral over $\ell$ is dominated by $\ell\gg1/\ell_{\text{AdS}}$, then for nonzero mode number $n$ we can neglect the constant term in the tidal tensor. Then the integral becomes 
\begin{align}
\int_{0}^{\infty}\frac{d\ell}{\ell^{1/2}}\frac{\exp\left(-\frac{\Delta \tau}{2}\left(\frac{1}{\ell}+\ell m^2\right)\right)}{\prod_{a=0}^{3}\sqrt{\det\left(-\ell^{-2}(\partial_\tau^2-{\Phi^a}_a)\right)}}\Gamma\left(1-\frac{3\pi ML}{8\ell r_+^4}\right)^2\Gamma\left(1+\frac{3\pi ML}{4\ell r_+^4}\right),\label{lintegral}
\end{align}
Note that the exponential factor has a saddle at $\ell=1/m$. We assume that the dimension of the operator does not scale with the string length, so that $m\ll \ell_{\text{AdS}}$ in string units. Then $\ell\gg 1/\ell_{\text{AdS}}$, consistent with our assumption above. Plugging in the saddle and renormalizing as in Section 3 gives
\begin{align}
\langle O(t,\phi)O(0,0)\rangle=\left(\frac{L}{r_+}\right)^{2m}\Gamma\left(1-\frac{3\pi mML}{8r_+^4}\right)^2\Gamma\left(1+\frac{3\pi mML}{4 r_+^4}\right).\label{correlatoranswer}
\end{align}
We have omitted the zero mode determinant in (\ref{lintegral}) because it gives a subleading power law in $L$ at large $m$. The zero mode was discussed in the pure AdS case in \cite{maxfield}.\\
\indent Finally, we may take the limit as $L\to -i\infty$ in (\ref{correlatoranswer}). We find
that the gamma functions are exponentially suppressed, and the singularity is resolved. In fact, (\ref{correlatoranswer}) vanishes in the limit. Note that this does not mean that the full correlator vanishes on the light cone, since there are other spacelike geodesics connecting the two boundary points. These other geodesics, which wind around the photon sphere, give the dominant (and finite) contribution to the correlation function on the light cone.  Though this addresses the
question raised in the introduction, it would be more illuminating if we were able
to compute the correlator near the singularity in physical kinematics as well as at the
singularity. 
This would require analyzing the fate of the negative eigenvalue mode in the full black hole geometry away from the Penrose region, and
new tools are likely needed for this purpose.  \section{Asymptotically flat black holes} \label{s8}
\begin{figure}[t]
        \center{\includegraphics[scale=.7]
        {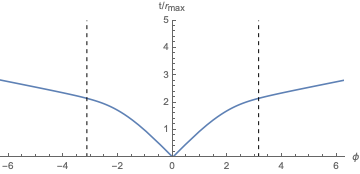}}
   \caption{The light cone in the asymptotically flat case is a single smooth curve. Here we set $r_s/r_{\text{max}}=1/10$. }\label{lightconeplotflat}
      \end{figure}

We can easily generalize the singularity to asymptotically flat black holes (see also \cite{schwarzschildsingularity}). Here we compute the two-point function at some large radius $r_{\text{max}}$, and the light cone will depend on $r_{\text{max}}$. The differential equations are now
\begin{align}
\frac{d\phi}{dr}&=\pm \frac{r_+r_-}{r_s\sqrt{(r^2-r_+^2)(r^2-r_-^2)}}\notag\\
\frac{dt}{dr}&=\pm \frac{1}{\sqrt{(1-(r_+/r)^2)(1-(r_-/r)^2)}}\frac{1}{1-(r_s/r)^2}\end{align}
where the turning points are 
\begin{align}
r_{\pm}=\sqrt{\frac{1\pm\sqrt{1-4(r_sE/L)^2}}{2(E/L)^2}}.
\end{align}
Integrating from $r$ to $r_{\text{max}}$ gives
\begin{align}
\Delta\phi&=-\frac{2r_-}{r_s}F\left(\text{ArcSin}\left(\frac{r_+}{r}\right),\frac{r_-^2}{r_+^2}\right)\big|^{r_{\text{max}}}_{r_+}\notag\\
\Delta t&=-2r_+\left(\frac{r_s^2}{r_+^2}\Pi\left(\frac{r_s^2}{r_+^2},\text{ArcSin}\left(\frac{r_+}{r}\right),\frac{r_-^2}{r_+^2}\right)+F\left(\text{ArcSin}\left(\frac{r_+}{r}\right),\frac{r_-^2}{r_+^2}\right)-E\left(\text{ArcSin}\left(\frac{r_+}{r}\right),\frac{r_-^2}{r_+^2}\right)\right)\notag\\
&\hspace{5 mm}+2r\sqrt{\left(1-\frac{r_-^2}{r^2}\right)\left(1-\frac{r_+^2}{r^2}\right)}\big|_{r_+}^{r_{\text{max}}}
.\end{align}
In the limit where $r_{\text{max}}$ goes to infinity, we get the late time behavior
\begin{align}
\Delta \phi &=\frac{2r_-}{r_s}K\left(\frac{r_-^2}{r_+^2}\right)\notag\\
\Delta t&=2r_{\text{max}}+2r_+\left(\frac{r_s^2}{r_+^2}\Pi\left(\frac{r_s^2}{r_+^2},\frac{r_-^2}{r_+^2}\right)+K\left(\frac{r_-^2}{r_+^2}\right)-E\left(\frac{r_-^2}{r_+^2}\right)\right).
\end{align}
\indent We see that the dependence on $r_{\text{max}}$ at late times is simple. It reflects the fact that for geodesics that get reasonably close to the black hole, the time it takes to get back out to infinity is of order $2r_{\text{max}}$ for large $r_{\text{max}}$. There is no analog of the geodesic hitting the boundary and bouncing back into the bulk. So the picture is a bit simpler, as shown in Figure \ref{lightconeplotflat}.
      \section{Future directions}\label{s9}
     In this paper we have shown how singularities that are present in the thermal two-point function at infinite $\lambda$ are resolved by bulk strings at finite $\lambda$. There are various possible extensions of this result. First, one could analyze the singularity structure of higher point functions. When the number of points is greater than three, there can be a bulk Landau diagram which leads to a boundary singularity. It would be interesting to understand the conditions for such singularities to be resolved.\\
  \indent Another interesting direction, in which we are currently investigating \cite{progress}, is to understand the generalization of the results here to Kerr black holes, corresponding to a CFT at finite temperature and rotation. In this case there is no longer a rotational symmetry on the boundary sphere, so the singularities become more complicated. In particular, for equatorial geodesics there are two photon radii, one for prograde and one for retrograde orbits. The prograde photon orbit approaches the horizon radius in the extremal limit, leading to the possibility of probing horizon-scale physics.\\
  \indent Finally, we have only discussed the bulk point of view in this work, but one could also try to understand these singularities from the CFT perspective. In particular, can the singularity be seen in the conformal bootstrap \cite{bootstrap} or in a prototypical CFT like maximally supersymmetric Yang-Mills theory? If so, it would be interesting to understand if $1/\lambda$ corrections can be resummed to resolve the singularity directly in CFT. 

\acknowledgments
We thank P. Di Vecchia,  V. Hubeny, T. Jacobson, E. Martinec, 
D. Meltzer, M. Mirbabayi, M. Rangamani, G. Sarosi, S. Shenker,  D. Stanford, 
E. Silverstein, and N. Warner for discussion. 
The work of H.O. is supported in part by
U.S.\ Department of Energy grant DE-SC0011632, by
the World Premier International Research Center Initiative,
MEXT, Japan, by JSPS Grant-in-Aid for Scientific Research 17K05407  and 20K03965,
and by JSPS Grant-in-Aid for Scientific Research on Innovative Areas
15H05895.
H.O. thanks the Aspen Center for Theoretical Physics, which is supported by
the National Science Foundation grant PHY-1607611,  where part of this work was done. The work of M.D. is supported by JSPS KAKENHI Grant Number 20K14465.

  \appendix
\section{Singularities in some limiting cases}\label{a1}
In this appendix we will explore several limits of parameter space where the singularity structure of the two-point function can be analyzed exactly. In all three cases the only singularity will be on the ordinary light cone.
\subsection{Infinite volume in 1+1 dimensions}
We are interested in a CFT on $S^{1}$ at finite temperature, in the limit where the radius of the circle becomes infinite. In this limit we have a CFT on a cylinder $\mathbf{R}\times S^1$, so the two-point function is completely determined by conformal invariance. It is
\begin{align}
\langle O(t,x)O(0,0)\rangle\sim \frac{1}{\left(\sinh^2\left(\frac{\pi(t+x)}{\beta}\right)\sinh^2\left(\frac{\pi(t-x)}{\beta}\right)\right)^{\Delta}}.
\end{align}
The only singularities of this function are at $t=\pm x$, which is the ordinary light cone. Therefore if there is a nontrivial singularity in the two-point function in 1+1 dimensions, then it must disappear at infinite volume.
\subsection{Free field theory}\label{freefield}
We consider a scalar field on $S^{d-1}$ at finite temperature, introducing a mass to deal with infrared divergences. For simplicity we take $d=3$. The mode expansion is 
\begin{align}
\Phi(\tau,\theta,\phi)=\sum_{\ell,m,k}e^{2\pi i k\tau/\beta}Y_\ell^m(\theta,\phi)\Phi_{k,\ell, m}
\end{align}
The Euclidean action is 
\begin{align}
\frac{1}{2}\sum_{k,\ell ,m}\left(\left(\frac{2\pi k}{\beta}\right)^2+\ell(\ell+1)+M^2\right)\Phi_{k,\ell,m}\Phi_{-k,\ell,-m}.
\end{align}
The two-point function on the equator is 
\begin{align}
\langle \Phi(\tau,\pi/2,\phi)\Phi(0,\pi/2,0)\rangle= \sum_{\ell,m,k}\frac{e^{2\pi i k\tau/\beta}}{\left(\frac{2\pi k}{\beta}\right)^2+\ell(\ell+1)+M^2}Y^m_{\ell}(\pi/2,\phi)Y^{-m}_\ell(\pi/2,0).
\end{align}
We can do the sum over $k$ using Matsubara techniques. We get 
\begin{align}
\sum_{\ell }\frac{(2\ell+1)P_\ell(\cos\phi)}{\sqrt{\ell(\ell+1)+M^2}}\left(\frac{e^{-\sqrt{\ell(\ell+1)+M^2}\tau}}{1-\exp\left(-\beta\sqrt{\ell(\ell+1)+M^2}\right)}-(\tau,\beta)\to (-\tau,-\beta)\right).
\end{align}
The divergences come from large $\ell$ in the sum. Expanding the Legendre polynomials at large $\ell$, we get singularities at $\tau=\pm i \phi$, which is just the ordinary light cone.
\subsection{Rational CFT in two dimensions}
Finally, we consider the finite-temperature two-point function of a rational CFT in two dimensions,
\begin{align}
Z(\tau, z; \bar{\tau}, \bar{z}) = {\rm Tr} \left( q^{L_0 - \frac{c}{24}} \bar{q}^{\bar{L}_0 - \frac{c}{24}}
O(z, \bar z) O(0, 0) \right),
\end{align}
with $q = e^{2 \pi i \tau}$ and the periodicities $z \sim z + 1 \sim z + \tau$. 
If the CFT is rational, we can express $Z$ as a finite sum over conformal blocks, 
\begin{align}
Z(\tau, z; \bar{\tau}, \bar{z})  = \sum_i F_i(\tau, z) \bar{F}_i(\bar{\tau}, \bar{z}) .
\label{ConformalBlockSum}
\end{align}
\indent We keep $(\tau, \bar{\tau})$ in the Euclidean domain (namely, $\bar \tau$ is the complex conjugate of $\tau$) 
and analytically continue in $(z, \bar z)$ to the Lorentzian domain. Thus, we are interested in studying properties of 
$Z(\tau, z; \bar{\tau}, \bar{z})$ as a function of two 
independent complex variables $z$ and $\bar z$ with fixed $(\tau, \bar \tau)$. Since each $F_i(\tau, z)$ is holomorphic in $z$,
it can only have singularities at points in the complex $z$-plane and not along a curve. Since the sum in (\ref{ConformalBlockSum}) is finite,
any singularities of $Z(\tau, z; \bar{\tau}, \bar{z})$ should also be of this type, namely either at a point in the $z$-plane or a point in 
the $\bar z$-plane. The red curves in Figure \ref{lightconeplot} are of this type, since they correspond to either $z=0$ or $\bar{z}=0$. On the other hand, 
the blue curves are not of this type since they are expressed in terms of an equation involving both $z$ and $\bar z$. For example, their 
asymptotic forms for large $t$ is 
\begin{align}
z + \bar z = \pm v_\gamma (z - \bar z),
\end{align}
with $v_\gamma$ defined in (\ref{PhotonSphereVelocity}).
 Singularities along such curves cannot arise from the finite sum
over the conformal blocks in  (\ref{ConformalBlockSum}).  Though singularities along the blue curves are also absent
in a semi-classical gravity in AdS$_3$, as we noted in Section 2, this argument gives yet another indication that the only singularity in a generic CFT is on the ordinary light cone.

\section{The phase of the determinants at early times}\label{phase}

\indent In this appendix we will evaluate the determinants at early times using the Gelfand-Yaglom theorem \cite{dunne}. Recall that this theorem first requires us to find a function $y$ satisfying 
\begin{align}
(-\partial_\tau^2+V(\tau))y(\tau)=0,\hspace{10 mm}y(\tau_i)=0,\hspace{10 mm}y'(\tau_i)=1.
\end{align}
Once we find such a $y$, we can evaluate the determinant as 
\begin{align}
\det(-\partial_\tau^2+V(\tau))=y(\tau_f).
\end{align}

\indent We will compute the determinant using the shockwave approximation. For example consider the $x^1$ direction. The function $y$ is then given by (\ref{x1ansatz}), with 
\begin{align}
a_1=-\frac{e^{in\tau_i}}{2in},\hspace{10 mm}a_1^\dagger=\frac{e^{-in\tau_i}}{2in}.
\end{align}
The determinant is then equal to
\begin{align}
y(\tau_f)=\frac{\sin(n(\tau_f-\tau_i))}{n}-\frac{3\pi p_vL   M}{2n^2r_+^4}\sin(n\tau_f)\sin(n\tau_i).
\end{align}
\indent We now take this answer and project it onto the vacuum. We get 
\begin{align}
\det\left(-\partial_\tau^2-n^2+p_v^2A_{11}(p_v\tau)\right)=\frac{e^{in(\tau_f-\tau_i)}}{2in}\left(1-\frac{3\pi ip_vL   M}{4nr_+^4}\right).
\end{align}
The first factor would be there in flat space, and is treated in \cite{gsw,gilesthorn}. Therefore the factor we are interested in is the second factor. Including the two attractive modes, the product of the determinants over $n$ is
\begin{align}
\prod_{n=1}^{\infty}\left(1-\frac{3\pi ip_vL   M}{4nr_+^4}\right)\left(1+\frac{3\pi ip_vL   M}{8nr_+^4}\right)^2=\frac{1}{\Gamma\left(1-\frac{3\pi ip_vL   M}{4r_+^4}\right)\Gamma\left(1+\frac{3\pi ip_vL   M}{8r_+^4}\right)^2}\label{detphase}.
\end{align}
Note that the magnitude of (\ref{detphase}) reproduces (\ref{detmag}), as promised.
\section{The tidal tensor for spacelike geodesics}\label{tidaltensor}
We consider a spacelike geodesic that is almost lightlike. We want to expand the metric around this geodesic so that we can analyze the worldsheet theory. We closely follow the analysis of \cite{marck}, although that reference analyzes timelike and not spacelike geodesics. \\
\indent The first step is to find an orthonormal tetrad $\lambda_a^\mu$ that is parallel transported along the geodesic. Once we get this tetrad we may define the tidal tensor as 
\begin{align}
\Phi_{ab}=\lambda_4^\mu\lambda_a^\nu\lambda_b^\rho\lambda_4^\sigma R_{\mu\nu\rho\sigma},
\end{align}
where $\lambda_4^\mu=\dot{x}^\mu$ is the tangent vector to the geodesic. The metric near the geodesic then takes the form 
\begin{align}
ds^2=\eta_{ab}dx^a\, dx^b+\left(1+\Phi_{ab}x^ax^b\right)\,(dx^4)^2.
\end{align}
The transverse indices $a,b$ now run from 0 to 3. There are other terms in the expansion of the metric, for instance terms proportional to $x^a\, x^b\, dx^c\, dx^d$. However these terms can be neglected. The reason is that the classical solution to the worldline equations of motion has $\dot{x}^4\not=0$. Therefore when we expand around this solution in the action, the term proportional to $\Phi_{ab}$ is quadratic in the $x^a$ fields. However $\dot{x}^a=0$, so $x^a\, x^b\, \dot{x}^c\, \dot{x}^d$ is actually quartic in the fields, not quadratic. Therefore it can be neglected.\\
\indent Now let us compute $\lambda_a^\mu$ for $a\not=1$. There are two obvious ones, 
\begin{align}
\lambda_2&=\frac{1}{r}\partial_\theta\notag\\
\lambda_3&=\frac{1}{r}\partial_\psi,
\end{align}
where the coordinates on the $S^3$ are defined by
\begin{align}
d\psi^2+\sin^2\psi(d\theta^2+\sin^2\theta d\phi^2).
\end{align}
The geodesic is at $\psi=\theta=\pi/2$, so these basis vectors are normalized correctly. \\
\indent What about the other two? The strategy of \cite{marck} is to first complete the orthonormal basis with particularly simple vectors $\tilde{\lambda}_0$ and $\tilde{\lambda}_1$ that are not parallel transported, and then solve the parallel transport equations by applying a time dependent rotation (or a boost in our case) on $\tilde{\lambda}_0$ and $\tilde{\lambda}_1$. So we make the ansatz 
\begin{align}
\tilde{\lambda}_0=\tilde{\lambda}_0^t\partial_t+\tilde{\lambda}_0^r\partial_r. 
\end{align}
The normalization condition and the dot product with $\lambda_4$ determine 
\begin{align}
\tilde{\lambda}_0^t&=\frac{\dot{r}}{\sqrt{1-L^2/r^2}(r^2+1-M/r^2)}\notag\\
\tilde{\lambda}_0^r&=\frac{E}{\sqrt{1-L^2/r^2}}.
\end{align}
Clearly this only works for $r>L$, so let us assume that for now. Similarly, we make the ansatz 
\begin{align}
\tilde{\lambda}_1=\tilde{\lambda}_2^t\partial_t+\tilde{\lambda}_2^r\partial_r+\tilde{\lambda}_2^\phi\partial_\phi
\end{align}
This completes the orthonormal basis if 
\begin{align}
\tilde{\lambda}_1^t&=\frac{E}{\sqrt{r^2/L^2-1}(r^2+1-M/r^2)}\notag\\
\tilde{\lambda}_1^r&=\frac{\dot{r}}{\sqrt{r^2/L^2-1}}\notag\\
\tilde{\lambda}_1^\phi&=-\frac{\sqrt{1-L^2/r^2}}{r}.
\end{align}
\indent As mentioned above, $\tilde{\lambda}_0$ and $\tilde{\lambda}_1$ are not yet parallel transported. Therefore we make the ansatz
\begin{align}
\lambda_0&=\tilde{\lambda}_0\cosh\eta+\tilde{\lambda}_1\sinh\eta\notag\\
\lambda_1&=\tilde{\lambda}_0\sinh\eta+\tilde{\lambda}_1\cosh\eta,
\end{align}
where $\eta$ is time-dependent. Now we need to solve the equations $\lambda^\mu_4\nabla_\mu \lambda_0=\lambda^\mu_4\nabla_\mu \lambda_1=0$. The covariant derivatives are best computed in Mathematica. We get 
\begin{align}
\dot{\eta}=\frac{EL}{r^2-L^2}.
\end{align}
\indent Finally we can compute the tidal tensor. We get 
\begin{align}
\Phi_{00}&=-\left(1+\frac{M}{r^4}+\frac{4M(L^2-r^2)\cosh^2\eta}{r^6}\right).\notag\\
\Phi_{01}&=\Phi_{10}=-\frac{2M(L^2-r^2)\sinh(2\eta)}{r^6}\notag\\
\Phi_{11}&=1+\frac{M}{r^4}-\frac{4M(L^2-r^2)\sinh^2\eta}{r^6}\notag\\
\Phi_{22}&=\Phi_{33}=1-\frac{M(2L^2-r^2)}{r^6}.
\end{align}
For $r<L$, we can do the same thing. We get 
\begin{align}
\Phi_{00}&=-\left(1+\frac{M}{r^4}-\frac{4M(L^2-r^2)\sinh^2\eta}{r^6}\right).\notag\\
\Phi_{01}&=\Phi_{10}=\frac{2M(L^2-r^2)\sinh(2\eta)}{r^6}\notag\\
\Phi_{11}&=1+\frac{M}{r^4}+\frac{4M(L^2-r^2)\cosh^2\eta}{r^6}\notag\\
\Phi_{22}&=\Phi_{33}=1-\frac{M(2L^2-r^2)}{r^6}.
\end{align}
\indent We are interested in almost null geodesics, so it's not hard to see that we can take $\eta=0$. The only terms that survive after taking large $L$ and $E$ for $r<L$ are 
\begin{align}\label{tidalequation}
\Phi_{00}&=-1\notag\\
\Phi_{11}&=1+\frac{4ML^2}{r^6}\notag\\
\Phi_{22}&=\Phi_{33}=1-\frac{2ML^2}{r^6}.
\end{align}
The constant terms come from the AdS curvature. Neglecting the constant terms at large $L$, we see that this matrix approaches the Penrose plane wave matrix, as expected.


\end{document}